\documentstyle[mprocl]{article}

\def\Journal#1#2#3#4{{#1} {\bf #2}, #3 (#4)}

\def\NCA{\em Nuovo Cimento}
\def\NIM{\em Nucl. Instrum. Methods}
\def\NIMA{{\em Nucl. Instrum. Methods} A}
\def\NPB{{\em Nucl. Phys.} B}
\def\PLB{{\em Phys. Lett.}  B}
\def\PRL{\em Phys. Rev. Lett.}
\def\PRD{{\em Phys. Rev.} D}
\def\ZPC{{\em Z. Phys.} C}
\def\st{\scriptstyle}
\def\sst{\scriptscriptstyle}
\def\mco{\multicolumn}
\def\epp{\epsilon^{\prime}}
\def\vep{\varepsilon}
\def\ra{\rightarrow}
\def\ppg{\pi^+\pi^-\gamma}
\def\vp{{\bf p}}
\def\ko{K^0}
\def\kb{\bar{K^0}}
\def\al{\alpha}
\def\ab{\bar{\alpha}}
\def\be{\begin{equation}}
\def\ee{\end{equation}}
\def\bea{\begin{eqnarray}}
\def\eea{\end{eqnarray}}
\def\CPbar{\hbox{{\rm CP}\hskip-1.80em{/}}}

\begin{document}

\title{REMARKS ON ANOMALOUS U(1) SYMMETRIES IN STRING THEORY
\footnote{Talk given at COSMO-99, International Workshop on Particle Physics
and the Early Universe, 27 September - 2 October 1999, ICTP, Trieste, Italy}}

\author{Hans Peter NILLES}

\address{Physikalisches Institut, Universit\"at Bonn,
Nussallee 12,
\\ D-53115 Bonn, Germany\\E-mail: nilles@th.physik.uni-bonn.de}


\maketitle\abstracts{
Anomalous U(1) gauge symmetries appear frequently 
both in heterotic and
type I/type II string theory. 
In the heterotic case we find at most a single 
anomalous U(1), while
in open string theories several such symmetries can appear.
We review the properties of the anomalous gauge bosons, 
the appearance of Fayet-Iliopoulos terms and 
the concept of heterotic-type I duality. 
Phenomenological applications of these symmetries
might be different  in the type I/type II case compared to the heterotic one.}
  
\def\Journal#1#2#3#4{{#1} {\bf #2}, #3 (#4)}

\def\NCA{\em Nuovo Cimento}
\def\NIM{\em Nucl. Instrum. Methods}
\def\NIMA{{\em Nucl. Instrum. Methods} A}
\def\NPB{{\em Nucl. Phys.} B}
\def\PLB{{\em Phys. Lett.}  B}
\def\PRL{\em Phys. Rev. Lett.}
\def\PRD{{\em Phys. Rev.} D}
\def\ZPC{{\em Z. Phys.} C}
\def\st{\scriptstyle}
\def\sst{\scriptscriptstyle}
\def\mco{\multicolumn}
\def\epp{\epsilon^{\prime}}
\def\vep{\varepsilon}
\def\ra{\rightarrow}
\def\ppg{\pi^+\pi^-\gamma}
\def\vp{{\bf p}}
\def\ko{K^0}
\def\kb{\bar{K^0}}
\def\al{\alpha}
\def\ab{\bar{\alpha}}
\def\be{\begin{equation}}
\def\ee{\end{equation}}
\def\bea{\begin{eqnarray}}
\def\eea{\end{eqnarray}}
\def\CPbar{\hbox{{\rm CP}\hskip-1.80em{/}}}

\def\be{\begin{equation}}
\def\ee{\end{equation}}
\def\ba{\begin{array}}
\def\ea{\end{array}}
\def\bea{\begin{eqnarray}}
\def\eea{\end{eqnarray}}
\def\GeV{{\rm GeV}}
\def\tr{{\rm tr}}
\def\thefootnote{\fnsymbol{footnote}}
\def\chib{{\bar\chi}}
\def\psib{{\bar\psi}}
\def\nn{\nonumber}

\def\wS{S}
\def\wT{T}                                                  .

\def\sS{{\cal S}}
\def\sT{{\cal T}}

\def\NPB#1#2#3{{Nucl.~Phys.} {{B#1}} (19#2) #3}
\def\PLB#1#2#3{{Phys.~Lett.} {{B#1}} (19#2) #3}
\def\PRD#1#2#3{{Phys.~Rev.} {{D#1}} (19#2) #3}




\newcommand{\barre}[1]{%
        \setbox1=\hbox{$#1$} \dimen2=\ht1 \dimen3=\dp1 \dimen4=\wd1
        \setbox2=\hbox{\sl /}
        \dimen1=\wd1 \advance\dimen1 by -\wd2 \divide\dimen1 by 2
        \advance\dimen1 by \wd2 \advance\dimen1 by 0.4pt
        \setbox3=\hbox to \wd1{\hss \box1 \kern -\dimen1 \box2\hss}
        \ht3=\dimen2 \dp3=\dimen3 \wd3=\dimen4
        \box3
        }


\section{Introduction}

The appearance of anomalous U(1) gauge symmetries in the framework
of string theory has received considerable attention. 
While primarily the motivation to study such symmetries was of theoretical
origin, it was soon realized that there could be interesting
applications to model building. This included the possible role of 
induced Fayet-Iliopoulos terms for gauge and supersymmetry breakdown and
the appearance of global symmetries relevant for
the strong CP-problem and questions of baryon and lepton number
conservation. Cosmological applications can be found in a 
discussion of D-term inflation and the creation of the cosmological  
baryon asymmetry. 

In string theory, anomalous U(1) gauge symmetries
can serve as tools to study detailed properties of duality
symmetries and the question of supersymmetry breakdown. Most recently
this became apparent in attempts to relate orbifold compactifications
of the perturbative heterotic string to orientifolds of Type II string
theory. In the present talk I shall report on results obtained in
collaboration with Z. Lalak and S. Lavignac. Lack of space and time
allows just a summary of basic results. For details and a more 
complete list of references we refer the reader to the original
publications \cite{LLN,LLN2}.


\section{Anomalous $U(1)$'s in heterotic string theory}

In field theoretic models we were taught to discard anomalous gauge
symmetries 
in order to avoid inconsistencies. This was even extended for the
condition on the trace of the charges $\sum_i Q_i = 0$ of a
$U(1)$ gauge symmetry because of mixed gauge and gravitational
anomalies \cite{Alvarez}. Moreover a nonvanishing trace of
the $U(1)$ charges would reintroduce quadratic divergencies
in supersymmetric theories through a one-loop Fayet-Iliopoulos
term \cite{Fischler}. In string theory we then learned that
one can tolerate anomalous $U(1)$ gauge symmetries due to the
appearance of the Green-Schwarz mechanism \cite{GS} that 
provides a mass for the anomalous gauge boson. In fact,
anomalous $U(1)$ gauge symmetries are common in string
theories and could be useful for various reasons. 

In the case of the heterotic string
one obtains models with at most one anomalous $U(1)$,
and the Green-Schwarz mechanism involves the so-called
model independent axion (the pseudoscalar of the dilaton
superfield $S$). The number of potentially anomalous gauge
bosons is in general limited by the number of antisymmetric tensor fields
in the ten-dimensional ($d=10$) string theory. This explains
the appearance of only one such gauge boson in the perturbative
heterotic string theory and leads to specific correlations between the
various (mixed) anomalies
\cite{Kobayashi}. This universal anomaly structure is tied
to the coupling of the dilaton multiplet to the various gauge
bosons.  

The appearance of a nonvanishing trace of the $U(1)$ charges leads
to the generation of a Fayet-Ilopoulos term $\xi^2$ at one loop.
In the low energy effective field theory this would be quadratically
divergent, but in string theory this divergence is cut off through
the inherent regularization due to modular invariance. One obtains
\cite{DSW,Atick}
\begin{equation}
  \xi^2\ \sim {1\over({S+S^*})}M_{\rm Planck}^2 \sim M_{\rm String}^2
\label{eq:FIterm}
\end{equation}
where $(S+S^*)\sim 1/g^2$ with the string coupling constant $g$.
The Fayet-Iliopoulos term of order of the string scale
$M_{\rm String}$ is thus generated in perturbation theory. 
This could in principle lead to a breakdown of supersymmetry,
but in all known cases there exists a supersymmetric minimum in
which charged scalar fields receive nonvanishing vacuum
expectation values (vevs), that break $U(1)_A$ (and even other
gauge groups) spontaneously. This then leads to a mixing of
the goldstone boson (as a member of a matter supermultiplet)
of this spontaneous breakdown and the
model-independent axion (as a member of the
dilaton multiplet) of the Green-Schwarz mechanism. 
One of the linear combinations will provide a mass to
the anomalous gauge boson. The other combination will obtain
a mass via nonperturbative effects that might even be 
related to an axion-solution of the strong CP-problem \cite{strong_CP}.
As we can see from (\ref{eq:FIterm}), both the mass of the
$U(1)_A$ gauge boson and the value of the Fayet-Iliopoulos term
$\xi$ are of the order of the string scale. Nonetheless,
models with 
an anomalous $U(1)$ have been considered under various
circumstances and lead to a number of desirable 
consequences. Among those are
the breakdown of some additional nonanomalous gauge groups
\cite{Font},
a mechanism to parametrize the fermion mass spectrum in
an economical way \cite{fermion_masses},
the possibility to induce a breakdown of supersymmetry
\cite{susy_breaking},
a satisfactory incorporation of D-term inflation \cite{inflation},
and the possibility for an axion solution of the strong
CP-problem \cite{strong_CP}.

The nice property of the perturbative heterotic string theory 
in the presence of an anomalous $U(1)$ is the 
fact that both $\xi$ and the mass of the anomalous gauge boson
are induced dynamically and not just put in by hand. Both of them,
though, are of order of the string scale $M_{\rm String}$, which
might be too high for some of the applications. 
We will now compare this for the case of
type I and type II orientifolds.

\section{Anomalous $U(1)$'s in type I and type II theories}

We consider $d=4$ string models of both open and
closed strings that are derived from either type I or type II
string theories in $d=10$ by appropriate orbifold or orientifold
projections \cite{orientifolds}. It was noticed, that in
these cases more than a single anomalous $U(1)$ symmetry could be obtained
\cite{Ibanez_orientifolds}. This lead to the belief that here we can
deal with a new playground of various sizes of $\xi$'s and 
gauge boson masses in the phenomenological applications.

The appearance of several anomalous
$U(1)$'s is a consequence of the fact that these models contain
various antisymmetric tensor fields in the higher dimensional
theory and the presence of a generalized Green-Schwarz mechanism
\cite{Sagnotti_generalized,Berkooz} involving axion fields in new 
supermultiplets $M$. In the
type II orientifolds under consideration these new axion fields 
correspond to twisted fields in the Ramond-Ramond sector of the
theory.

From experience with the heterotic case it was then assumed
\cite{Kakushadze_Z_3} that for each anomalous $U(1)$ a 
Fayet-Iliopoulos term was induced dynamically. With a mixing of
the superfields $M$ and the dilaton superfield $S$ one 
hoped for $U(1)_A$ gauge boson masses of various sizes in 
connection with various sizes of the $\xi$'s.

The picture of duality between heterotic orbifolds and 
type II orientifolds as postulated in \cite{Sagnotti}
seemed to work even in the presence
of several anomalous $U(1)$
gauge bosons assuming the presence of Fayet-Iliopoulos terms
in perturbation theory and the presence of the generalized
Green-Schwarz mechanism. So superficially everything seemed to
be understood. But apparently the situation turned out to
be more interesting than anticipated.

\section{Some Surprises}

There appeared two decisive results that
initiated renewed interest in these questions and
forced us to reanalyse this situation \cite{LLN}. 
The first one concerns the inspection of the anomaly
cancellation mechanism in various type II orientifolds.
As was observed by Ib\'a\~nez, Rabadan and Uranga \cite{Ibanez_anomalous},
in this class of models there is no mixing between the dilaton
multiplet and the $M$-fields. It is solely the latter
that contribute to the anomaly cancellation. Thus the
dilaton that is at the origin of the Green-Schwarz mechanism
in the heterotic theory does not participate in that
mechanism in the dual orientifold picture.
The second new result concerns the appearance of
the Fayet-Iliopoulos terms. As was shown by Poppitz \cite{Poppitz}
in a specific model,
there were no $\xi$'s generated in one-loop perturbation theory.
The one loop contribution vanishes
because of tadpole cancellation in the given theory. 
This result seems to be of more general validity and
could have been anticipated from general arguments,
since in type I theory a (one-loop) contribution to
a Fayet-Iliopoulos term either vanishes or is quadratically
divergent, and the latter divergence is avoided by the
requirement of tadpole cancellation. Of course, there is
a possibility to have tree level contributions to the $\xi$'s,
but they are undetermined, in contrast to the heterotic case
where $\xi$ is necessarily nonzero because of the
one loop contribution. In type II theory such a contribution
would have to be of nonperturbative origin.
In the heterotic theory the mass of the anomalous gauge boson was
proportional to the value of $\xi$. If a similar result would hold
in the orientifold picture, this would mean that some of
the $U(1)$ gauge bosons could become arbitrary light or even
massless, a situation somewhat unexpected from our experience
with consistent quantum field theories. In any case, a careful
reevaluation of several questions is necessary in the
light of this new situation. Among those are:
the size of the $\xi$'s,
the size of the masses of anomalous $U(1)$ gauge bosons,
the relation of $\xi$ and gauge boson mass,
as well as the fate of heterotic - type IIB orientifold duality,
which we will discuss in the remainder of this talk.

The questions concerning the anomalous gauge boson masses have been
answered in \cite{LLN}. Generically they are large, of order of the
string scale, even if the corresponding Fayet-Iliopoulos terms vanish.
This is in agreement with the field theoretic expectation that
the masses of anomalous gauge bosons cannot be small or even zero.
There is one possible exception, however. In the limit that
gauge coupling constant tends to zero, one could have vanishing
masses. In this case, one would deal with a global  U(1)
that can be tolerated in field theory even if it is anomalous.


\section{Heterotic-Type I Duality}

Models containing anomalous $U(1)$ factors offer an
arena to study details of Type I/II - Heterotic duality in
four dimensions. This duality,
is of the weak coupling - strong coupling type in ten
dimensions. In four dimensions the relation between the
heterotic and type I dilatons is
\begin{equation}
\phi_H = \frac{1}{2} \phi_I - \frac{1}{8} \log (G_I)
\end{equation}
where $G_I$ is the determinant of the metric of the compact 6d
space, which depends on  moduli fields. For certain relations
between the dilaton and these moduli fields we thus have a duality
in four dimensions which maps a weakly coupled theory to another
weakly coupled theory.

For the remainder of the discussion we have to be very careful
with the definition of heterotic - type I duality. 
Such a duality has
first been discussed in \cite{Polchinski} in ten dimensions.
It was explicitely understood as a duality between the
original $SO(32)$ type I theory and the heterotic theory with the same
gauge group, that is a duality between two theories that both have
one antisymmetric tensor field in ten dimensions. This is a very well
established duality symmetry which will not be the focus of our 
discussion here. We would like to concentrate on a 
four dimensional duality symmetry
between more general type II orientifolds and the heterotic $SO(32)$
theory first discussed in \cite{Sagnotti}. We call this heterotic -
type II orientifold  duality. It would relate theories that have a 
different number of antisymmetric tensor field in their ten
dimensional origin.

The pairs of models which we study are type IIB orientifolds models in 4d
and their candidate heterotic duals which can be found in the existing
literature
\cite{Sagnotti,Kakushadze_Z_3,Kakushadze_Z_7,Kakushadze_Z_3xZ_3,Ibanez_Z_3,Ibanez_orientifolds,Kakushadze_orientifolds,Lykken}.
As an example consider the $Z_3$ orientifold/orbifold \cite{Sagnotti}.
The type IIB
orientifold model has the gauge group $G=SU(12) \times SO(8)
\times U(1)_A$ where the $U(1)_A$ factor is anomalous. The
anomalies are non-universal and get cancelled by means of the
generalized Green-Schwarz mechanism. This mechanism involves twenty-seven
twisted singlets $M_{\alpha \beta \gamma}$, a particular
combination of which combines with the anomalous vector
superfield to form a massive multiplet. After the
decoupling of this heavy vector multiplet we obtain the
nonanomalous model with the gauge goup $G'=SU(12) \times SO(8)$.

On the heterotic side, with the heterotic $SO(32)$ superstring
compactified on the orbifold $T^6/Z_3$, the gauge group is $G=SU(12)
\times SO(8) \times U(1)_A$ and the $U(1)_A$ is again anomalous.
Its anomalies, however, are universal in this case, and a
universal, only dilaton-dependent, Fayet-Iliopoulos term is
generated. In this case there are also fields which are charged only
under the anomalous $U(1)$  that can compensate for the Fayet-Iliopoulos
term by assuming a nontrivial vacuum
 expectation value, without breaking the gauge group
any further; a combination of these fields and of the dilaton supermultiplet
is absorbed by the anomalous vector multiplet. These nonabelian singlets
are the counterparts of the
$M_{\alpha \beta \gamma}$ moduli of the orientifold model. However, on
the heterotic side we have additional states charged under $U(1)_A$ (and
also under $SO(8)$) the counterparts of which are not present in
the orientifold model. These unwanted states become heavy in a supersymmetric
manner through the superpotential couplings \cite{Kakushadze_Z_3}.
Below the scale of the heavy gauge boson mass we have a pair of
models whose spectra  fulfil the duality criteria.

There are, however, arguments that this duality symmetry might
not be universally valid. The first doubts came from a study of
the $Z_7$ examples in \cite{LLN}. There it was shown that the spectra
of the two candidate duals did not match for certain values of the
moduli fields. These doubts were confirmed in a calculation of
gauge coupling constants \cite{ABD}. Finally it was shown that
certain global symmetries \cite{IbanezT,Scrucca}
 that were found to hold on the heterotic side did
not have counterparts in the orientifold picture \cite{LLN2}.

\section{Outlook}

The presence of anomalous $U(1)$ symmetries can have interesting
phenomenological applications both in the heterotic and the type I case.
In heterotic string compactifications, the presence of an anomalous $U(1)$
shows up primarily in the existance of a nonvanishing
Fayet-Iliopoulos term $\xi$. If such a term is somewhat smaller than the
Planck scale this could explain
the origin and
hierarchies of the small dimensionless parameters in the low-energy
lagrangian, such as the Yukawa couplings \cite{fermion_masses}, in terms of
the ratio $\, {\xi}/M_{Pl}\,$. 
In explicit string models, $\xi$ is found to be of the order of magnitude
necessary to account for the value of the Cabibbo angle. Furthermore,
the universality of the mixed gauge anomalies implies a successful
relation between the value of the weak mixing angle at unification and the
observed fermion mass hierarchies \cite{weak_angle}.
The anomalous $U(1)$ could also play an 
important role in supersymmetry breaking:
not only does it take part in its mediation from the hidden sector to the
observable sector (as implied by the universal Green-Schwarz relation among
mixed gauge anomalies), but also it can trigger the breaking of
supersymmetry itself, due to an interplay between the anomalous $D$-term
\cite{susy_breaking}
and gaugino condensation \cite{HPN}. It would be interesting to look
at this questions in the framework of the heterotic $E_8\times E_8$
M-theory \cite{HoravaWitten} in the presence of anomalous $U(1)$
symmetries, generalizing previous results of supersymmetry
breakdown \cite{NOY}.
Cosmologically, the presence of an anomalous $U(1)$ might have
important applications in the discussion of inflationary models:
in particular
its Fayet-Iliopoulos term can dominate the vacuum energy of the early Universe,
leading to so-called  
D-term inflation \cite{inflation}. Finally, the heterotic anomalous $U(1)$
might be at the origin of a solution of
the strong CP problem \cite{strong_CP}, while providing an
acceptable dark matter candidate.
Since there is no exact heterotic - type II orientifold duality
one may now ask whether the anomalous $U(1)$'s 
present in type IIB orientifolds
are likely to have similar consequences - or even have the potential to solve
some of the problems encountered in the heterotic case. Certainly, the
implications will differ somewhat.
In the heterotic case, the phenomenological
implications of the  $U(1)_X$ rely on the appearance of a
Fayet-Iliopoulos term whose value, a few orders of magnitude below the Planck
mass, is fixed by the anomaly. The situation is different in 
in the orientifold case,
where the Fayet-Iliopoulos terms are moduli-dependent.
The freedom that is gained by the possible adjustment of the
Fayet-Iliopoulos term allows, for example, to cure the 
problems of $D$-term inflation in
heterotic models \cite{Halyo}, where $\xi$ turned out to be too
large.
This possible choice of $\xi$ is payed for by a loss of predictivity.
In that respect, one may conclude that the orientifold anomalous $U(1)$'s are
not that different from anomaly-free $U(1)$'s, whose Fayet-Iliopoulos terms
are unconstrained and can be chosen at will. This might also question
the possible use of these $U(1)$'s for an axion solution of the
strong CP-problem.
Still, these anomalous
U(1) symmetries might play an important role in phenomenological
applications.

\section*{\bf Acknowledgements}

I would like to thank Z. Lalak and S. Lavignac for interesting
discussions and collaboration.
This work was partially supported by the European Commission programs
ERBFMRX-CT96-0045 and CT96-0090.




\section*{References}
\begin{thebibliography}{Ref}

\newcommand{\bi}[1]{\bibitem{#1}}


\bibitem{LLN} Z. Lalak, S. Lavignac and H.P. Nilles, hep-th/9903160;
Nucl. Phys. B559 (1999) 48

\bibitem{LLN2} Z. Lalak, S. Lavignac and H.P. Nilles, hep-th/9912206 

\bibitem{Alvarez} L. Alvarez-Gaume and E. Witten, Nucl. Phys. B234 (1984) 269.

\bibitem{Fischler} W. Fischler, H.P. Nilles, J. Polchinski, S. Raby and 
L. Susskind, Phys. Rev. Lett. 47 (1981) 757.

\bibitem{GS} M. Green and J. Schwarz, Phys. Lett. B149 (1984) 117.

\bibitem{Kobayashi} T. Kobayashi and H. Nakano, Nucl. Phys. B496 (1997) 103.

\bibitem{DSW} M. Dine, N. Seiberg and E. Witten, Nucl. Phys. B289 (1987) 317.

\bibitem{Atick} J. Atick, L. Dixon and A. Sen, Nucl. Phys. B292 (1987) 109;
M. Dine, I. Ichinose and N. Seiberg, Nucl. Phys. B293 (1987) 253.

\bibitem{strong_CP} J.E. Kim, Phys. Lett. B207 (1988) 434;
E.J. Chun, J.E. Kim and H.P. Nilles, Nucl. Phys. B370 (1992) 105;
H. Georgi, J.E. Kim and H.P. Nilles, Phys. Lett. B437 (1998) 325.

\bibitem{Font} A. Font, L.E. Ibanez, H.P. Nilles and F. Quevedo,
Nucl. Phys. B307 (1988) 109; {\it ibid.}, B310 (1988) 764.

\bibitem{fermion_masses} L. Ib\'a\~nez, G. G. Ross, Phys. Lett. B332 (1994) 
100; P. Bin\'etruy, P. Ramond, Phys. Lett. B350 (1995) 49;
V. Jain, R. Shrock, Phys. Lett. B352 (1995) 83;
E. Dudas, S. Pokorski, C.A. Savoy, Phys. Lett. B356 (1995) 45;
P. Bin\'etruy, S. Lavignac,  and P. Ramond,  Nucl. Phys. B477 (1996) 353;

\bibitem{susy_breaking} P. Bin\'etruy and E. Dudas, Phys. Lett. B389 (1996)
503; G. Dvali and A. Pomarol, Phys. Rev. Lett. 77 (1996) 3728;
Z. Lalak, Nucl. Phys. B521 (1998) 37;
N. Arkani-Hamed, M. Dine, S. P. Martin, Phys. Lett. B431 (1998) 329;
T. Barreiro, B. de Carlos, J.A. Casas, J.M. Moreno,
Phys. Lett. B445 (1998) 82.

\bibitem{inflation} J.A. Casas and C.  Mu\~noz, Phys. Lett. B216 (1989) 37;
J.A. Casas, J.M. Moreno, C. Mu\~noz and M. Quiros, Nucl. Phys. B328 (1989) 272;
P. Bin\'etruy and G. Dvali, Phys. Lett. B388 (1996) 241;
E. Halyo, Phys. Lett. B387 (1996) 43.

\bi{HPN} H. P. Nilles, \PLB {115}{82}{193}; Int. J. Mod. Phys. {A5} (1990) 4199

\bibitem{orientifolds} A. Sagnotti, in Cargese `87, ``Non-Perturbative
Quantum Field Theory'', eds. G. Mack et al. (Pergamon Press, Oxford, 1988),
p. 251;
P. Horava, Nucl. Phys. B327 (1989) 461, Phys. Lett. B231 (1989) 251.

\bibitem{Ibanez_orientifolds} G. Aldazabal, A. Font, L.E. Ibanez,
G. Violero, Nucl. Phys. B536 (1998) 29.

\bibitem{Sagnotti_generalized} A. Sagnotti, Phys. Lett. B294 (1992) 196.

\bibitem{Berkooz} M. Berkooz, R.G. Leigh, J. Polchinski, J.H. Schwarz,
N. Seiberg and E. Witten, Nucl. Phys. B475 (1996) 115.

\bibitem{Kakushadze_Z_3} Z. Kakushadze, Nucl. Phys. B512 (1998) 221.

\bibitem{Sagnotti} C. Angelantonj, M. Bianchi, G. Pradisi, A. Sagnotti, and
Ya.S. Stanev, Phys. Lett. B385 (1996) 96.

\bibitem{Ibanez_anomalous} L.E. Ibanez, R. Rabadan and A.M. Uranga,
Nucl. Phys. B542 (1999) 112

\bibitem{Poppitz} E. Poppitz, Nucl. Phys. B542 (1999) 31


\bibitem{Polchinski} J. Polchinski and E. Witten, Nucl. Phys. B460 (1996) 525.

\bibitem{HoravaWitten} P. Horava and E. Witten, Nucl. Phys. B460 (1996) 506;
Nucl. Phys. B475 (1996) 94

\bi{NOY} H. P. Nilles, M. Olechowski and M. Yamaguchi,
 \PLB{415}{97}{24}; \NPB{484}{97}{33}; \NPB{530}{98}{43}; 

\bibitem{Ibanez_aspects}  L.E. Ibanez, C. Munoz and S. Rigolin, 
Nucl. Phys. B553 (1999) 43

\bibitem{Kakushadze_Z_7} Z. Kakushadze and G. Shiu, Phys. Rev. D56 (1997) 3686.

\bibitem{Kakushadze_Z_3xZ_3} Z. Kakushadze and G. Shiu,
Nucl. Phys. B520 (1998) 75.

\bibitem{Ibanez_Z_3} L.E. Ibanez, JHEP 9807 (1998) 002.

\bibitem{Kakushadze_orientifolds} Z. Kakushadze, G. Shiu and S.-H.H. Tye,
Nucl. Phys. B533 (1998) 25.

\bibitem{Lykken} J. Lykken, E. Poppitz and S.P. Trivedi, 
Nucl. Phys. B543 (1999) 105

\bibitem{ABD} I. Antoniadis, C. Bachas and E. Dudas, 
Nucl. Phys. B560 (1999) 93

\bibitem{IbanezT} L. E. Ibanez, R. Rabadan and A.M. Uranga, hep-th/9905098

\bibitem{Scrucca} C. A. Scrucca and M. Serone, hep-th/9912108

\bibitem{Cvetic} M. Cvetic, L. Everett, P. Langacker and J. Wang,
JHEP 9904:020,1999

\bibitem{weak_angle} L. Ib\'a\~nez, Phys. Lett. B303 (1993) 55;
P. Bin\'etruy and P. Ramond, Phys. Lett. B350 (1995) 49.

\bibitem{Halyo} E. Halyo, Phys. Lett. B454 (1999) 223 


\end{thebibliography}
\end{document}